\documentclass[11pt]{article}
\usepackage{hyperref}
\pdfoutput=1
\begin{document}
\title{Straight-line and turning locomotion of {\it Paramecia} }
\author{Saikat Jana$^1$, Matthew Giarra$^2$, Pavlos Vlachos$^2$ and Sunghwan Jung$^1$ \\
\\\vspace{6pt} $^1$Department of Engineering Science and Mechanics
\\\vspace{6pt} $^2$Department of Mechanical Engineering, \\ Virginia Tech, Blacksburg, VA 24061, USA}

\maketitle
\begin{abstract}
In this fluid dynamics video we investigate the flow field around straight-line swimming and right and left turning {\it Paramecia} using micro- particle image velocimetry ($\mu$PIV). A {\it Paramecium} controls its ciliary beating to produce different fluid velocities on either side of its body. This phenomenon is visualized by applying $\mu$PIV to images in which {\it Paramecia} swim in a dilute suspension of 1 $\mu$m polystyrene spheres. {\it Paramecia} that swim straight exhibit similar magnitudes of velocity on either side of their bodies. In contrast, right-turning {\it Paramecia} exhibit greater magnitudes of velocity on their right sides, while left-turning organisms show the opposite. 

 \end{abstract}
\end{document}